# Bragg-Grating Enhanced Narrow-Band Spontaneous Parametric Down Conversion

Li Yan,[1,*] Lijun Ma,[2] and Xiao Tang[2]

[1]*Department of Computer Science and Electrical Engineering, University of Maryland, Baltimore County, 1000 Hilltop Circle, Maryland 21250, USA*
[2]*Information Technology Laboratory, National Institute of Standards and Technology, 100 Bureau Drive, Gaithersburg, Maryland 20899, USA*
[*]*liyan@umbc.edu*

**Abstract:** We propose a new method to narrow the linewidth of entangled photons generated from spontaneous parametric down conversion (SPDC). An internal Bragg grating is incorporated onto a nonlinear optical crystal waveguide. We study theoretically the spectral characteristics of SPDC under two Bragg grating structures. We show that using the Bragg grating with a midway π-phase shifter, it is a promising way to generate narrow-line entangled photons.

## 1. Introduction

Quantum information science and technology open up a fascinating future for the research, development and application in information technologies. Quantum communication is one of its most important and practical applications. For highly efficient long distance quantum communication, quantum repeater is a necessary component to recover the fidelity of the quantum information encoded on single photons after long transmission. According to the DLCZ protocol [1], a quantum repeater is formed mainly by an entangled photon source for teleporting long distance and a quantum memory for storing of quantum information. In quantum memory, an atomic ensemble is used to store quantum information carried by single photons. The linewidth of the entangled photons should be comparable to the atomic transition linewidth of a few MHz.

The current mainstream method to generate entangled photons is through a nonlinear optical process called spontaneous parametric down conversion (SPDC) [2]. For easy system integration with a low power pump laser, the nonlinear optical crystal is made into a waveguide, and crystal's polarity is spatially periodically poled, resulting in quasi-phase matching. However, the entangled photons from such periodically poled nonlinear optical crystal waveguide have relatively broad line-width, typically several nanometers. Therefore, reduction of the bandwidth of entangled photons by SPDC is important for quantum teleportation with quantum memory. A straight forward way of narrowing the SPDC bandwidth is through passive filtering [3,4]. This approach suffers the disadvantage of additional loss of the desired signal idler photons. Below- threshold optical parametric oscillator is another approach [5-11]. Although spectral width at specific signal or idler wavelength can be quite narrow, because of the closeness of the resonator longitudinal modes, multiple signal and idler lines exist, and it has bulky and complex configuration.

In this paper, we propose a new method to narrow the linewidth of entangled photons by SPDC. A quasi-phase-matched nonlinear optical crystal waveguide is incorporated with an internal Bragg grating. This approach, if implemented, has only a single longitudinal mode, and its configuration will be much compact. We study theoretically the spectral characteristics of SPDC incorporated with internal Bragg grating under finite quasi-phase-matching. Two grating structures are considered. We show that using a Bragg grating with a midway π-phase shifter, it is a promising way to generate narrow-line entangled photons.

## 2. Model

The schematic structure of the proposed SPDC with internal Bragg grating is shown in Fig. 1. A Bragg grating is written onto a periodically poled nonlinear optical crystal waveguide. We consider two Bragg grating structures. One has a continuous grating modulation. Another has a half grating period of flat spacer at the middle of the full Bragg grating, which is commonly used in semiconductor distributed feed-backed lasers. Note that Implemention of a Bragg grating structure on a periodically poleld nonliner crystal waveguide is technically practical, and a similar technology (with a continuous grating) has been demonstrated for distgributed feedback optical parametric oscillation [12] and enhanced second harmonic generation [13].

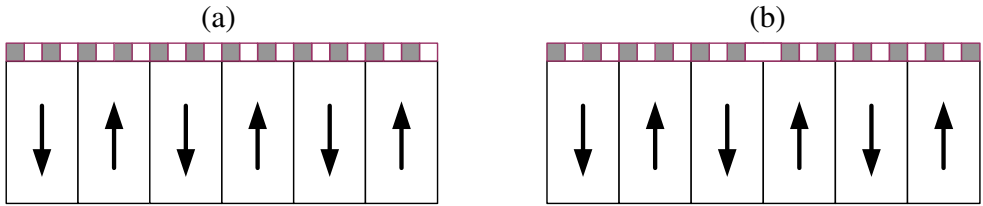

Fig. 1. Periodically poled nonlinear optical crystal waveguide with Bragg grating. (a) Continuous grating; (b) grating with midway π-phase shift.

We proceed with the study of optical parametric conversion under the influence of internal Bragg grating by the classical method, i.e., we treat the signal, idler and pump as classical waves, with equivalent noise input fields for signal and idler. For SPDC, the pump field is virtually non-depleted and we treat it as constant. Thus one considers the coupling evolution of the signal and the idler waves. Including the forward and backward traveling waves, the signal and idler fields take the form

$$E_s(z,t) = \mathrm{Re}[\tilde{E}_s(z)\exp(j\omega_s t)] \tag{1}$$

$$E_i(z,t) = \mathrm{Re}[\tilde{E}_i(z)\exp(j\omega_i t)] \tag{2}$$

$$\tilde{E}_s(z) = A_s(z)\exp(-j\beta_s z) + B_s \exp(j\beta_s z) \tag{3}$$

$$\tilde{E}_i(z) = A_i(z)\exp(-j\beta_i z) + B_i \exp(j\beta_i z) \tag{4}$$

where the $A$'s are the forward waves and $B$'s are the backward waves. The wave equations for the signal and idler fields are

$$\left[\frac{d^2}{dz^2} + \omega_s^2 \mu \varepsilon_s\right] \tilde{E}_s = -\omega_s^2 \mu \left(\tilde{P}_{G,s} + \tilde{P}_{NL,s}\right) \quad (5)$$

$$\left[\frac{d^2}{dz^2} + \omega_i^2 \mu \varepsilon_i\right] \tilde{E}_i = -\omega_i^2 \mu \left(\tilde{P}_{G,i} + \tilde{P}_{NL,i}\right) \quad (6)$$

For simplicity we assume the dielectric constant to vary sinusoidal in z-direction. For the continuous grating as in Fig. 1(a), we write the modulation of susceptibility as

$$\Delta\varepsilon(z) = \varepsilon_g \cos\beta_G z \quad (7)$$

We assume a non-degenerate optical parametric down conversion and that the Bragg grating couples strongly only the signal wave, i.e. the Bragg condition is satisfied only for the signal $2\beta_s \approx \beta_G$. Therefore we neglect the backward pump and idler waves. The Bragg grating perturbation polarization is given by

$$\tilde{P}_{G,s}(z) = \Delta\varepsilon(z)\left[A_s(z)\exp(-j\beta_s z) + B_s(z)\exp(j\beta_s z)\right] \quad (8)$$

and $\tilde{P}_{G,i} \cong 0$. The nonlinear polarizations for the signal and idler waves are given by, respectively

$$\tilde{P}_{NL,s}(z) = 2d(z)A_i^* A_p \exp[-j(\beta_p - \beta_i)z] \quad (9)$$

$$\tilde{P}_{NL,i}(z) = 2d(z)A_s^* A_p \exp[-j(\beta_p - \beta_s)z] \quad (10)$$

Also for simplicity, let the periodically poled nonlinearity be written as

$$d(z) = d_Q \cos\beta_Q z \quad (11)$$

The quasi-phase matching condition is $\beta_p - \beta_s - \beta_i \approx \beta_Q$. Under the slowly-varying-envelop approximation and keeping the nearly phase-matched and grating-resonant terms, the signal-grating-coupled parametric down conversion process is governed by the following coupled differentially equations:

$$\frac{dA_s}{dz} = -j\kappa_g B_s \exp(j\Delta\beta_G z) - j\kappa_d A_i^* \exp(-j\Delta\beta_Q z) \quad (12)$$

$$\frac{dA_i^*}{dz} = j\kappa_d^* A_s \exp(-j\Delta\beta_Q z) \quad (13)$$

$$\frac{dB_s}{dz} = j\kappa_g A_s \exp(-j\beta_G z) \quad (14)$$

where

$$\kappa_g = \frac{\mu\varepsilon_g \omega_s c}{4n_s} \quad (15)$$

$$\kappa_d = \sqrt{\frac{\mu\omega_s \omega_i}{\varepsilon_0 n_s n_i}} \frac{d_Q \tilde{E}_p(0)}{4} \quad (16)$$

$$\Delta \beta_G = 2\beta_s - \beta_G \qquad (17)$$

$$\Delta \beta_Q = \beta_p - \beta_s - \beta_i - \beta_Q \qquad (18)$$

In the above equations, $A_s$ is the forward signal field, $A_i$ is the forward idler field, and $B_s$ is the backward signal field. The field amplitudes are normalized, but keeping the same symbols, such that the modular square of the amplitude gives the photon flux density in a particular wave. $\Delta \beta_G$ is a measure of frequency detuning away from the Bragg grating line center, and $\Delta \beta_Q$ is the phase mismatch away from the signal/idler line-center. When the Bragg grating line center and the quasi-phase-matched signal line center line up, $\Delta \beta_G$ and $\Delta \beta_Q$ are related by the difference of group indexes of the signal and idler waves

$$\Delta \beta_Q = [(n_i - n_s)/2n_s]\Delta \beta_G \qquad (19)$$

When dispersion is neglected, the group indexes can be approximated by the corresponding refractive indexes. For the second Bragg grating structure, it is equivalently to have a π-phase shift at the midpoint of the full sinusoidal Bragg grating

$$\Delta \varepsilon(z) = \begin{cases} \varepsilon_g \cos \beta_G z & 0 \leq z < L/2 \\ -\varepsilon_g \cos \beta_G z & L/2 \leq z \leq L \end{cases} \qquad (20)$$

corresponding to $\kappa_g$ changing a sign in the second half region.

Note that distributed feedback optical parametric amplification (OPA) and oscillation with a continuous Bragg grating have been studied by the classical theory, in which phase matching among pump, signal and idler waves is assumed and for OPA only the signal wave has input [14]. Although the equations presented above are similar to that in [14], the calculation in [14] does not properly describe the SPDC process. In SPDC, both signal and idler photons are generated from vacuum fluctuations. Accurate calculation must be based on the quantum theory. With definite input signal and idler fields, Eqs. (12)-(14) describe a coherent process, which is sensitive to the relative phases of the input signal and idler fields. Note that there is no first-order correlation between the signal and idler photons generated from SPDC. In the classical theory approximation, the equivalent input fields of signal and idler must have random phases. We take equivalent input fields for the forward signal and idler waves with random phases at $z = 0$ and zero input field for the backward signal wave at $z = L$. The physical SPDC response is ensemble-averaged over random noise inputs.

## 3. Results

As a baseline, Fig. 2 shows the transmission and reflection characteristics of the two Bragg grating structures for the forward and backward signal waves, with grating coupling parameter $\kappa_g L = 3$, a reasonable grating coupling strength [14]. The unit of frequency detuning is $\Delta \beta_G L/2$. For the continuous Bragg grating, around zero detuning, the signal wave is mostly reflected with a broadband. In contrast, for the Bragg grating structure with a midway π-phase shift, the transmitted signal wave has a sharp peak at zero detuning.

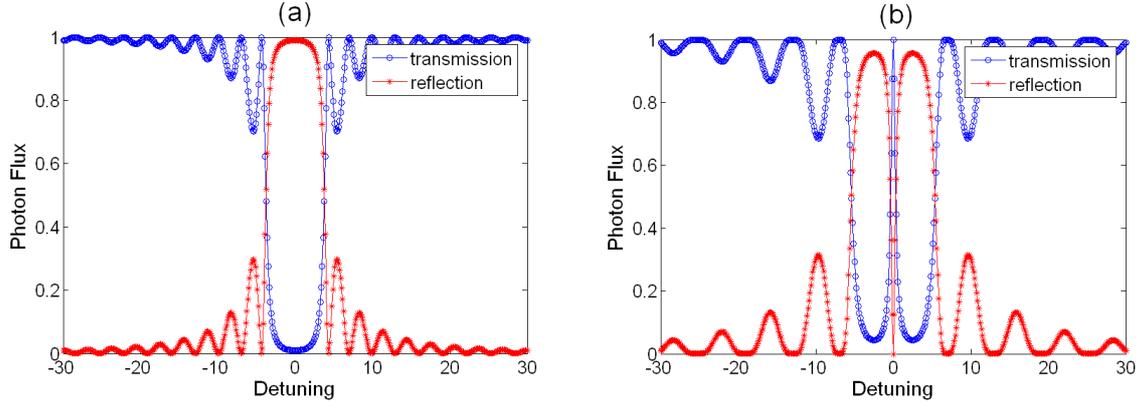

Fig. 2. Transmission and reflection characteristics of two Bragg grating structures.
(a) Continuous grating; (b) grating with midway π-phase shift.

Fig. 3 shows the spectral characteristics spontaneous optical parametric down conversion with two types of Bragg grating in resonance with the signal wave. As a comparison, the broadband spectral response of parametric down conversion is also shown. For illustration, we take $\Delta\beta_Q = 0.1\Delta\beta_G$ and $\kappa_d L = 1.1$. The calculated SPDC bandwidth ($\Delta\beta_G L / 2 \approx 8\pi$) is about 0.3 nm for a nonlinear optical crystal waveguide length of 1 cm, a nominal wavelength of 1.5 μm and refractive index of 2.2. The refractive index difference of commonly used nonlinear optical crystals is about one-order of magnitude smaller, corresponding to an SPDC bandwidth of a few nm for 1-cm length of nonlinear crystal waveguide [15]. With an internal continuous Bragg grating ($\kappa_g L = 3$, $\kappa_d L = 1.1$), within the stop-band of the Bragg grating, most of the signal flux is in the backward direction, while the idler's flux is also reduced. However, the signal and idler photons are generally still broadband. In contrast, with the Bragg grating structure that has a midway π-phase shift ($\kappa_g L = 3$, $\kappa_d L = 0.7$), the forward signal and idler photons peark sharply, with comparable high fluxes, at zero detuning. The bandwidth is about one order of magnitude narrower than the stop-band of the Bragg grating. With a stronger grating coupling strength, the spetral narrowing will be more prominent.

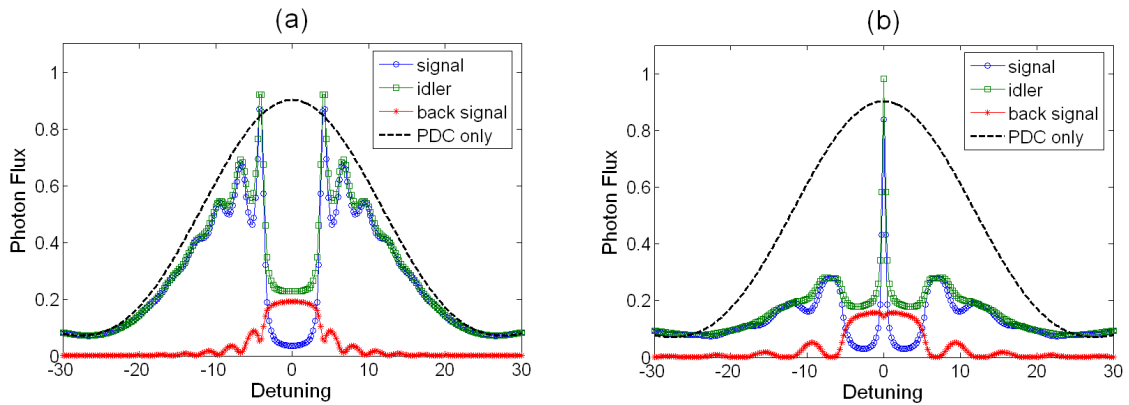

Fig. 3. Signal and idler photon flux densities with and without a Bragg grating. (a) Continuous grating; (b) grating with π-phase shift.

The narrow-band peak of PDC is reminascent of a spectral resonant characteristics. In the context of quantum nature of SPDC, the resonant characteristics should be understood as a result of the imposition of a constraint on the nonlinear system by the midway π-phase-shifted Bragg grating and consequently forcing the spontaneous parametric down conversion into only a narrow band. In fact, the grating structure with a midway flat spacer can be viewed physically as a mircro cavity of $\lambda_s/4$ length and bounded by two Bragg reflectors. This internal-grating-enhanced SPDC is in contrast with passive filtering by an external Bragg grating, as evidenced by the fact that the spectral peak can reach a given flux level under a lower pump level compared to the necessary pump level for PDC without the Bragg grating.

**4. Summary**

We propose a new approach to narrow down linewidth of entangled photons generated from spontaneous parametric down conversion by incorporating a Bragg grating structure with a midway π-phase shifter onto a nonlinear optical crystal waveguide. We theoritically study the linewidths of signal and idler photons from the SPDC process and show that the approach is promising to generate very narrow linewidth signal and idler photons, an entangled photon source very usefull for quantum information and communication applications.